\documentstyle[11pt]{article}
\begin{document}

\title{\hfill {\small IFUM-918-FT \bigskip\bigskip\bigskip}\\
{\bf Universal Kounterterms in Lovelock AdS
gravity\footnote{Presented at 3rd RTN Workshop: Constituents,
Fundamental Forces and Symmetries of the Universe, Valencia, Spain ,
1-5 Oct 2007.}}}
\author{Georgios Kofinas $^{a}$ and Rodrigo Olea $^b$\medskip \\
{\small {\em $^a$ Department of Physics and Institute of Plasma
Physics, University of Crete, 71003 Heraklion, Greece}}\\
{\small {\em $^{b}$ INFN, Sezione di Milano, Via Celoria 16,
I-20133,
Milano, Italy.}}\\
{\small {\tt gkofin@phys.uoa.gr, rodrigo.olea@mi.infn.it}}}
\date{}
\maketitle

\begin{abstract}
We show the universal form of the boundary term (Kounterterm series)
which regularizes the Euclidean action and background-independent
definition of conserved quantities for any Lovelock gravity theory
with AdS asymptotics (including Einstein-Hilbert and
Einstein-Gauss-Bonnet). We discuss on the connection of this
procedure to the existence of topological invariants and
Chern-Simons forms in the corresponding dimensions.

\end{abstract}

\section{Introduction}

Lovelock gravity \cite{lovelock} is the natural generalization of
two basic features of General Relativity: general covariance and (at
most) second-order field equations. It is also known to be free of
ghosts when expanded on a flat space \cite{ghost-free}.

The introduction of (negative) cosmological constant leads to
solutions that are not asymptotically flat. Therefore, the Euclidean
bulk action and the conserved quantities evaluated on these
solutions contain divergences that should be cured using a suitable
regularization mechanism. The AdS/CFT correspondence provides a
regularizing prescription for Einstein-Hilbert gravity coming from
the holographic reconstruction of asymptotically AdS (AAdS)
spacetimes in Fefferman-Graham coordinates \cite{fg,henniskende}.
This procedure results into the addition to the bulk (plus the
corresponding
Gibbons-Hawking term) of an intrinsic (Dirichlet) counterterms series ${\cal %
L}_{ct}$ \cite{bala,EJM}. It is important to stress, however, that
due to the increasing complexity of the algorithm to construct the
counterterms in higher dimensions, holographic renormalization has
been unable to provide a full expression for ${\cal L}_{ct}$ in any
dimension. Moreover, as it can be seen in Einstein-Gauss-Bonnet
gravity, the asymptotic resolution of the equations of motion in
terms of a given data at the boundary turns out to be extremely
difficult. These arguments motivate the search for another
regularization scheme for Lovelock AdS gravity.

An alternative prescription for boundary terms which achieve the
finiteness of both Euclidean action and conserved quantities in
Einstein-Hilbert gravity with AdS asymptotics has been provided in
\cite{OleaJHEP, OleaKounter} as given polynomials in the extrinsic
and intrinsic curvatures (Kounterterms). As the natural extension of
what happens in Einstein-Gauss-Bonnet gravity \cite{Kofinas-Olea},
we show below that, whenever the effective cosmological constant is
negative, the form of the Kounterterms is universal for all gravity
theories of the Lovelock type \cite{Kofinas-Olea-Lovelock}.

%%%%%%%%%%%%%%%%%%%%%%%%%%%%%%

\section{Kounterterms in Einstein-Hilbert AdS gravity}
\label{EH}

AdS gravity action in $D=d+1$ dimensions can be regularized using
Kounterterms $B_{d}$, which are boundary terms with explicit
dependence on
the extrinsic curvature\footnote{%
Here, hatted curvatures refer to the ones of the bulk manifold. The
cosmological constant is $\Lambda =-(D-1)(D-2)/2\ell^{2}$ in terms
of the
AdS radius $\ell $.}%
\begin{equation}
I=\frac{1}{16\pi G}\,\int\limits_{M}d^{d+1}x\,\sqrt{-g}\left( \hat{R}%
-2\Lambda \right) +c_{d}^{EH}\,\int\limits_{\partial M}d^{d}x\,B_{d}
\label{I}
\end{equation}%
where $c_{d}^{EH}$ is a given coupling constant, fixed demanding a
well-defined variational principle for AAdS spacetimes.

One of the simplest examples of the use of Kounterterms
regularization is given by four-dimensional AdS gravity. Due to the
fact that the Gauss-Bonnet term in four dimensions is a topological
invariant, one has always the
freedom to add it to the bulk action%
\begin{equation}
I_{4}=\int\limits_{M}d^{4}x\,\sqrt{-g}\left[ \frac{1}{16\pi G}\,\left( \hat{%
R}-2\Lambda \right) +\alpha \,(\hat{R}^{\mu \nu \sigma \rho
}\hat{R}_{\mu \nu \sigma \rho }-4\hat{R}^{\mu \nu }\hat{R}_{\mu \nu
}+\hat{R}^{2})\right] , \label{I4+GB}
\end{equation}%
with an arbitrary coupling constant $\alpha $. The presence of the
Euler-Gauss-Bonnet term ${\cal E}_{4}$ does not modify the bulk
dynamics.

In any gravity theory, the Wald's entropy formula \cite{wald}
amounts to taking derivatives of the bulk action with respect to the
Riemann tensor and suitably projecting this quantity at the horizon.
This method provides the correct value of the entropy for a large
number of different gravity actions. The direct application of this
prescription to Eq.(\ref{I4+GB}) shifts the entropy by a constant
$32\pi ^{2}\alpha \chi (\Sigma _{2})$, where $\chi (\Sigma _{2})$\
stands for the Euler characteristic of the two-dimensional
transversal section.

However, an arbitrary coupling of the Euler term is inconsistent
from the point of view of the variational principle. Indeed, the
theory has a well-posed variational principle for AAdS spacetimes
which satisfy the condition
\begin{equation}
\hat{R}_{\mu \nu }^{\kappa \lambda}+\frac{1}{\ell^{2}}\delta
_{\lbrack \mu \nu ]}^{[\kappa \lambda]}=0 \label{AAdSR}
\end{equation}
only if the coupling constant is chosen as $\alpha =\ell ^{2}/(64\pi
G)$ \cite{four}. This simple argument produces the on-shell
cancelation of divergences in the conserved quantities defined
through the Noether theorem.

On the other hand, for an arbitrary value of $\alpha $, the
Gauss-Bonnet invariant introduces additional divergent terms in the
Euclidean action, e.g., for Schwarzschild-AdS black holes. The
problem of a finite Euclidean action might be regarded as
disconnected from the variational principle. This is what makes
remarkable the fact that fixing $\alpha $ as above also provides a
mechanism to regularize the Euclidean action for asymptotically
AdS spacetimes and to reproduce the correct black hole thermodynamics \cite%
{OleaJHEP}.

Then, we see that the ambiguity present in Eq.(\ref{I4+GB}) is
removed by considering either the variational principle or the
regularization problem in what might be called {\em topological
}regularization of AdS gravity. However, the entropy is still
modified by a constant which, in the case of topological static
black holes is given by $vol(\Sigma_{2}^{k}) \ell ^{2}k/4G$, in
terms of the volume of the two-dimensional cross-section
$\Sigma_{2}^{k}$, which can be spherical, locally-flat or hyperbolic
(for $k=+1 ,0 ,-1$, respectively). Most of the thermodynamic
quantities are insensitive to an additive constant in the entropy,
but for $k=-1$ the entropy itself becomes negative for physical
black holes ($r_{+}<\ell $).

Negative entropy for hyperbolic black holes can be avoided
considering, instead, the boundary term $B_{3}$ that is locally
equivalent to the
Gauss-Bonnet term, which is dictated by the four-dimensional Euler theorem%
\begin{equation}
\int\limits_{M_{4}}d^{4}x~{\cal E}_{4}=32\pi ^{2}\chi
(M_{4})+\int\limits_{\partial M_{4}}d^{3}x\,B_{3}\,,  \label{ET4}
\end{equation}%
where $\chi (M_{4})$ is the Euler characteristic of the
four-dimensional manifold.

In Gauss-normal coordinates to describe the spacetime%
\begin{equation}
ds^{2}=g_{\mu \nu}dx^{\mu}dx^{\nu}=N^{2}(\rho )d\rho
^{2}+h_{ij}(x,\rho )dx^{i}dx^{j}, \label{GC}
\end{equation}%
the boundary $\partial M$ is defined by setting $\rho =const$. The term $%
B_{3}$ is a well-kown object known as second Chern form, which is
constructed using the second fundamental form (closely related to
the
extrinsic curvature), and whose explicit form in the coordinate frame (\ref%
{GC}) is%
\begin{equation}
B_{3}=4\sqrt{-h}\,\delta _{\lbrack
i_{1}i_{2}i_{3}]}^{[j_{1}j_{2}j_{3}]}\,K_{j_{1}}^{i_{1}}\left( \frac{1}{2}%
\,R_{j_{2}j_{3}}^{i_{2}i_{3}}(h)-\frac{1}{3}%
K_{j_{2}}^{i_{2}}K_{j_{3}}^{i_{3}}\right) \,.  \label{B3}
\end{equation}%
The above expression is a compact way of writing down a polynomial
in the
intrinsic curvature $R_{ij}^{kl}(h)$ associated to the boundary metric $%
h_{ij}$ and the extrinsic curvature
$K_{ij}=-\frac{1}{2N}\frac{\partial h_{ij}}{\partial \rho}$, using
the totally antisymmetrized Kronecker delta. Because the AdS action
can be regularized using Eq.(\ref{B3}) that does not depend only on
intrinsic tensors on $\partial M$, what we have is an alternative
counterterm series \cite{OleaJHEP}.

The generalization to higher even dimensions ($D=2n$) considers also
the regularizing effect of Euler term in the corresponding
dimension, which is not longer quadratic in the curvature for $D>4$
\cite{even}. The boundary
formulation equivalent to the bulk topological invariants is given by the $n$%
-th Chern form $B_{2n-1}$ which appears as the correction due to the
boundary in the $2n$-dimensional Euler theorem%
\begin{eqnarray}
B_{2n-1} &=&2n\sqrt{-h}\int\limits_{0}^{1}dt\,\delta _{\lbrack
i_{1}\cdots
i_{2n-1}]}^{[j_{1}\cdots j_{2n-1}]}\,K_{j_{1}}^{i_{1}}\left( \frac{1}{2}%
\,R_{j_{2}j_{3}}^{i_{2}i_{3}}-t^{2}K_{j_{2}}^{i_{2}}K_{j_{3}}^{i_{3}}\right)
\times  \nonumber \\
&&\,\,\,\,\,\,\,\,\,\,\cdots \times \left( \frac{1}{2}%
\,R_{j_{2n-2}j_{2n-1}}^{i_{2n-2}i_{2n-1}}-t^{2}K_{j_{2n-2}}^{i_{2n-2}}K_{j_{2n-1}}^{i_{2n-1}}\right)
\,.  \label{B2n-1}
\end{eqnarray}%

Then, the action (\ref{I}) becomes finite if we adjust the
coefficient of
the Kounterterms as%
\begin{equation}
c_{2n-1}^{EH}=\frac{1}{16\pi G}\frac{\left( -1\right) ^{n}\ell ^{2n-2}}{%
n\left( 2n-2\right) !}\,.
\end{equation}%
In the even-dimensional case one can also understand the
Kounterterms as a transgression form (a gauge-invariant extension of
a Chern-Simons density) for the Lorentz group $SO(2n-1,1)$\
\cite{eguchi}.

The absence of topological invariants of the Euler class in odd
dimensions can make quite difficult the problem of finding a
suitable Kounterterms series in this case. However, one is able to
get some insight from the simplest case and the fact that
three-dimensional AdS gravity can be formulated as a Chern-Simons
density for the AdS group. Indeed, using a single copy of the
Chern-Simons term for the group $SO(2,2)$, ones reproduces
Einstein-Hilbert
bulk action plus a boundary term%
\begin{equation}
I=Tr(AdA+\frac{2}{3}A^{3})=\frac{1}{16\pi G}\,\int\limits_{M}d^{3}x\,\sqrt{%
-g}\left( \hat{R}-2\Lambda \right) -\frac{1}{16\pi
G}\,\int\limits_{\partial M}d^{d+1}x\,\sqrt{-h}K\,,
\end{equation}%
which turns out to be half of the Gibbons-Hawking term. A well-posed
action principle is achieved by imposing a given boundary condition
on the extrinsic curvature, which is consistent to fixing the
conformal data for the metric at the boundary \cite{miol}. In doing
so, a single extrinsic boundary term is enough for the variational
and the regularization problems. It can also be shown that this
prescription is equivalent to the intrinsic regularization developed
by Balasubramanian and Kraus \cite{bala} up to a two-dimensional
topological invariant.

The generalization to higher odd-dimensional Einstein-Hilbert AdS
gravity of the extrinsic regularization mentioned above is far from
straightforward. However, an interesting hint is given by higher
dimensional Chern-Simons forms for the AdS group, which produce a
bulk gravity action of the Lovelock type. By simply restoring
Lorentz covariance at the boundary, one obtains a regularizing
boundary term constructed with the second fundamental form
\cite{Mora-Olea-Troncoso-Zanelli-CS,trans}.

The Kounterterm series in $D=2n+1$ dimensions is written in a
compact form
thanks to a double parametric integration%
\begin{eqnarray}
B_{2n}
&=&2n\sqrt{-h}\int\limits_{0}^{1}dt\int\limits_{0}^{t}ds\,\delta
_{\lbrack i_{1}\cdots i_{2n}]}^{[j_{1}\cdots
j_{2n}]}\,K_{j_{1}}^{i_{1}}\delta _{j_{2}}^{i_{2}}\left( \frac{1}{2}%
\,R_{j_{3}j_{4}}^{i_{3}i_{4}}-t^{2}K_{j_{3}}^{i_{3}}K_{j_{4}}^{i_{4}}+\frac{%
s^{2}}{\ell ^{2}}\,\delta _{j_{3}}^{i_{3}}\delta
_{j_{4}}^{i_{4}}\right)
\times  \nonumber \\
&&\,\,\,\,\,\,\,\,\,\cdots \times \left( \frac{1}{2}%
\,R_{j_{2n-1}j_{2n}}^{i_{2n-1}i_{2n}}-t^{2}K_{j_{2n-1}}^{i_{2n-1}}K_{j_{2n}}^{i_{2n}}+%
\frac{s^{2}}{\ell ^{2}}\,\delta _{j_{2n-1}}^{i_{2n-1}}\delta
_{j_{2n}}^{i_{2n}}\right) \,,  \label{B_2n}
\end{eqnarray}%
that, when written as a polynomial of the extrinsic and intrinsic
curvatures, produces the relative coefficients of the terms in the
expansion. The variational principle for AAdS spacetimes uniquely
determine
the weight factor of $B_{2n}$ as%
\begin{equation}
c_{2n}^{EH}=\frac{1}{16\pi G}\frac{\left( -1\right) ^{n}\ell ^{2n-2}}{%
2^{2n-2}n\left( n-1\right) !^{2}}\,,
\end{equation}%
value that could have also been calculated by means of the
cancelation of the highest-order divergences in the evaluation of
the Euclidean action. Remarkably, all the subleading divergent
contributions are also eliminated by the suitable choice of a single
coupling constant \cite{OleaKounter}.

%%%%%%%%%%%%%%%%%%%%%%%%%%%%%%

\section{Kounterterms in Einstein-Gauss-Bonnet AdS gravity}
\label{EGB}

In higher dimensions than four, the Gauss-Bonnet term is not longer
topological and thus, it contributes to the equations of motion. In
fact,
the field equations derived from the action%
\begin{equation}
I=\frac{1}{16\pi G} \int\limits_{M}d^{d+1}x\,\sqrt{-g}\left[  \hat{R}%
-2\Lambda +\alpha \,(\hat{R}^{\mu \nu \sigma \rho }\hat{R}_{\mu \nu
\sigma \rho }-4\hat{R}^{\mu \nu }\hat{R}_{\mu \nu
}+\hat{R}^{2})\right] +c_{d}^{EGB}\,\int\limits_{M}d^{d}x\,B_{d},
\label{EGB}
\end{equation}%
tell us that there are vacuum solutions which correspond to
constant-curvature spacetimes with an effective AdS radius
$\ell_{\!e\!f\!\!f}$ modified by the Gauss-Bonnet coupling as
\begin{equation}
\frac{1}{\ell _{\!e\!f\!\!f}^{2}}=\frac{1\pm \sqrt{1-4(D-3)(D-4)\alpha /\ell ^{2}}}{%
2(D-3)(D-4)\alpha }\,.  \label{elleffGB}
\end{equation}%
It was shown in Ref.\cite{Kofinas-Olea} that the regularization
prescription given above for Einstein-Hilbert gravity with AdS
asymptotics is still valid for the action (\ref{EGB}). This means
that the form of the Kounterterms in even and odd dimensions (Eqs.
(\ref{B2n-1}) and (\ref{B_2n}), respectively) is preserved, as long
as we change the AdS radius from $\ell $ to the effective one $\ell
_{\!e\!f\!\!f}$. The corresponding coupling constants are
modified by the variational principle as%
\begin{equation}
c_{2n-1}^{EGB}=\frac{1}{16\pi G}\frac{\left( -1\right) ^{n}\ell _{\!e\!f\!\!f}^{2n-2}%
}{n\left( 2n-2\right) !}\left( 1-\frac{2\alpha }{\ell _{\!e\!f\!f}^{2}}%
(D-2)(D-3)\right) \,,
\end{equation}%
for $D=2n$ and%
\begin{equation}
c_{2n}^{EGB}=\frac{1}{16\pi G}\frac{\left( -1\right) ^{n}\ell _{\!e\!f\!\!f}^{2n-2}}{%
2^{2n-2}n\left( n-1\right) !^{2}}\left( 1-\frac{2\alpha }{\ell _{\!e\!f\!\!f}^{2}}%
(D-2)(D-3)\right) \,,
\end{equation}%
for $D=2n+1$.

 For Einstein-Gauss-Bonnet gravity, the Dirichlet counterterms have been
 found only in five dimensions, which are constructed requiring general covariance
and not using holographic renormalization techniques for AAdS
spacetimes \cite{cvetic}. In turn, Kounterterms method provides a
prescription which leads to a finite value for both conserved
charges and Euclidean action in a background-independent fashion for
all dimensions. A covariant formula for the vacuum energy for AAdS
spacetimes has been derived, whose evaluation for topological static
black holes generalize the corresponding result in Einstein-Hilbert
gravity \cite{EJM}.

%%%%%%%%%%%%%%%%%%%%%%%%%%%%%%%%%%%%%%%%%%%%%%%%%%%%%%%%%%%%%%%%%%%%%%
\section{Kounterterms in Lovelock AdS gravity}
\label{Lovelock}

In $D=d+1$ dimensions, we will consider a regularized Lovelock
action of the form
\begin{equation}
\!\!\!\!\!\!\!\!\!\!\!\!\!\!\!I=\frac{1}{16\pi G}\int\limits_{M}\!%
\sum_{p=0}^{[(\!D-1\!)\!/2]}\alpha
_{p}L_{p}+c_{d}\int\limits_{\partial M}d^{d}x\,B_{d},
\label{lovelock}
\end{equation}%
where $G$ is the gravitational constant in $D$ dimensions, $L_{p}$
corresponds to the dimensional continuation of the Euler term in
$2p$ dimensions
\begin{equation}
\!\!\!\!\!\!\!\!\!\!\!\!\!\!\!L_{p}=\frac{1}{2^{p}}\sqrt{-g}\,\delta
_{\left[
\mu _{1}\cdots \mu _{2p}\right] }^{\left[ \nu _{1}\cdots \nu _{2p}\right] }\,%
\hat{R}_{\nu _{1}\nu _{2}}^{\mu _{1}\mu _{2}}\cdots \hat{R}_{\nu
_{2p-1}\nu _{2p}}^{\mu _{2p-1}\mu _{2p}}\,d^{D}x\,, \label{Lptensor}
\end{equation}%
and $\{\alpha _{p}\}$ is a set of arbitrary coefficients.

In Riemannian gravity, the equation of motion for a generic Lovelock
gravity is obtained varying with respect to the metric and takes the
form
\begin{equation}
E_{\mu }^{\nu }=\sum_{p=0}^{[(\!D-1\!)\!/2]}\frac{\alpha _{p}}{2^{p}}%
\,\,\delta _{\left[ \mu \mu _{1}\cdots \mu _{2p}\right] }^{\left[
\nu \nu _{1}\cdots \nu _{2p}\right] }\,\hat{R}_{\nu _{1}\nu
_{2}}^{\mu _{1}\mu _{2}}\cdots \hat{R}_{\nu _{2p-1}\nu _{2p}}^{\mu
_{2p-1}\mu _{2p}}=0. \label{EOMlovelock}
\end{equation}

For a given set of coefficients $\{\alpha _{p}\}$, the vacua of a
Lovelock theory are defined as the maximally symmetric spacetimes
that are globally of constant curvature. We will assume that all the
corresponding cosmological constants are real and negative, with
different effective AdS radii ${\ell _{\!e\!f\!\!f}}$ given by the
solutions to the equation
\begin{equation}
\sum\limits_{p=0}^{[(D-1)/2]}\!\,\frac{\alpha _{p}}{(D-2p-1)!}(-\ell
_{\!e\!f\!\!f}^{-2})^{p}=0.  \label{elleff}
\end{equation}

If we consider spacetimes which approach asymptotically to any of
these constant-curvature vacuum solutions, we can prove that the
Kounterterms regularization is universal, as the explicit expression
for the series keeps the form of Eqs. (\ref{B2n-1}) and (\ref{B_2n})
(in even and odd dimensions, respectively) for any Lovelock AdS
theory \cite{Kofinas-Olea-Lovelock}. Once again, we
only have to pass from the original AdS radius $\ell $ to the effective one $%
\ell _{\!e\!f\!\!f}$ given above. The finiteness of the Euclidean
action and Noether charges is achieved if and only if the coupling
constant are consistently
adjusted as%
\begin{equation}
c_{2n-1}=-\frac{1}{16\pi nG}\sum_{p=1}^{n-1}\frac{(-1)^{n-p}p\,\alpha _{p}%
}{(2n-2p)!}\ell _{\!e\!f\!\!f}^{2(n-p)},
\end{equation}%
and%
\begin{equation}
c_{2n}=-\frac{1}{16\pi G}\,\frac{(2n-1)!!}{2^{n-1}n!}\sum_{p=1}^{n}%
\frac{(-1)^{n-p}p\,\alpha _{p}}{(2n-2p+1)!}\ell
_{\!e\!f\!\!f}^{2(n-p)},
\end{equation}%
for even and odd (bulk) dimensions, respectively. As the
Kounterterms provides a method to get rid of all the divergences in
a background-independent way for Lovelock AdS gravity, it is natural
the appearance of a vacuum energy in $D=2n+1$ dimensions for
globally AdS spacetime, whose form generalizes the result of
Einstein-Hilbert and Einstein-Gauss-Bonnet theories.

\section{Conclusions}

We have argued that the regularization of any gravity theory of the
Lovelock type can be carried out using boundary terms which are a
given polynomial in the extrinsic and intrinsic curvatures
(Kounterterms), whose form is universal. Indeed, the only difference
appears at the level of the dimensionality, because even and odd
dimensions must be treated separately. In this context, this fact is
simply related to the existence of topological invariants of the
Euler class in even dimensions. This difference is not so
surprising, because in standard holographic renormalization there
appear technical differences between these two cases (existence of
vacuum energy and Weyl anomaly).

In the Dirichlet problem for gravity, one has to supplement the bulk
action with the Gibbons-Hawking term such that the variations of the
extrinsic curvature are canceled out. In Lovelock gravity, the Myers
procedure leads to generalized Gibbons-Hawking terms for the same
purpose \cite{Myers}. In this way, one understands why standard
counterterms can be only constructed up with covariant quantities of
the metric $h_{ij}$. However, the holographic
interpretation of a regularized stress tensor associates it to a metric $%
g_{(0)ij}$ at $\partial M$ which, on the contrary to $h_{ij}$, is
regular at the boundary.

In the Fefferman-Graham frame, the leading order of the expansion in
the
extrinsic curvature $K_{ij}$ is the same as the one of the boundary metric $%
h_{ij}$. This seems to explain why fixing the extrinsic curvature
instead of the metric one can still produce a finite action
principle for AdS gravity \cite{miol}.

A direct comparison between intrinsic counterterms and the
Kounterterms prescription has been given in Ref. \cite{mis-oleDCG}
for two particular Lovelock theories which feature a
symmetry-enhancement from Lorentz to AdS group, and where the first
problem is exactly solvable \cite{BOT}. This indicates that an
explicit comparison of both procedures might be possible also in
Einstein-Hilbert and Einstein-Gauss-Bonnet theories with AdS
asymptotics.

%%%%%%%%%%%%%%%%%%%%%%%%%%%%%%%%%%%%%%%%%%%%%%%%%%%%%%%%%%%%%%%%%%%%%%%%%%%%%%

\section*{Acknowledgements}
  G.K. is supported in part by EU grants
MRTN-CT-2004-512194 and MIF1-CT-2005-021982. R.O. is supported by
INFN. We thank G. Barnich, G. Compere, S. Detournay, J. Edelstein,
D. Klemm, O. Mi\v{s}kovi\'{c}, P. Mora and K. Skenderis for useful
conversations.

%%%%%%%%%%%%%%%%%%%%%%%%%%%%%%%%%%%%%%%%%%%%%%%%%%%%%%%%%%%%%%%%%%%%%%%%%%%%%%%%%%%
%%%%%%%%%%%%%%%%%%%%%%%%%%%%%%%%%%%%%%%%%%%%%%%%%%%%%%%%%%%%%%%%%%%%%%%%%%%%%%%%%%%
%%%%%%%%%%%%%%%%%%%%%%%%%%%%%%%%%%%%%%%%%%%%%%%%%%%%%%%%%%%%%%%%%%%%%%%%%%%%%%%%%%%

\end{document}